\documentclass[aaspp4,11pt]{aastex}
\eqsecnum
\def\beq{\begin{equation}}
\def\eeq{\end{equation}}
\def\ref{\reference}
\def\simge{\mathrel{%
   \rlap{\raise 0.511ex \hbox{$>$}}{\lower 0.511ex \hbox{$\sim$}}}}
\def\simle{\mathrel{
   \rlap{\raise 0.511ex \hbox{$<$}}{\lower 0.511ex \hbox{$\sim$}}}}
\begin{document}

\title{ 
LONG-TERM X-RAY VARIABILITIES OF THE SEYFERT
GALAXY MCG-2-58-22 : SECULAR FLUX DECREASE AND FLARES}

\author{Chul-Sung Choi}
\affil{Korea Astronomy Observatory, 61-1 Hwaam, Yusong, Taejon 305-348, Korea;
cschoi@kao.re.kr.}

\author{Tadayasu Dotani}
\affil{Institute of Space and Astronautical Science, 3-1-1 Yoshinodai,
Sagamihara, Kanagawa 229-8510, Japan; dotani@astro.isas.ac.jp.}

\author{Heon-Young Chang and Insu Yi}
\affil{Korea Institute for Advanced Study, 207-43 Cheongryangri, Dongdaemun,
Seoul 130-012, Korea;
hyc@kias.re.kr and iyi@kias.re.kr.}

\begin{abstract}
We have studied the long-term X-ray light curve (2$-$10 keV) 
of the luminous Seyfert 1 galaxy 
MCG-2-58-22 by compiling data, from various X-ray satellites, 
which together cover more than 20 years. We have found two distinct 
types of time variations in the light curve.
One is a gradual and secular decrease of the X-ray flux, 
and the other is the episodic increase of X-ray flux (or flare)
by a factor of 2$-$4 compared with the level expected from
the secular variation.
We detected 3 such flares in total; a representative
duration for the flares is $\sim $2 years, with intervening quiescent
intervals lasting $\sim\! 6-8$ years.
We discuss a few possible origins for these variabilities.
Though a standard disk instability theory may explain the displayed
time variability in the X-ray light curve, the subsequent accretions
of stellar debris, from a tidal disruption event caused by a
supermassive black hole in MCG-2-58-22, cannot be ruled out as 
an alternative explanation.

\end{abstract}


\section{INTRODUCTION}

X-ray observations of active galactic nuclei (AGN) show
that many of them are variable, over a range of amplitudes,
and on many timescales (Lawrence et al.\ 1985; Grandi et al.\ 1992; 
Mushotzky, Done, \& Pounds 1993; Nandra et al.\ 1997; Ulrich, Marachi, 
\& Urry 1997; Ptak et al.\ 1998; Turner et al.\ 1999). 
Variability of the X-ray emission is a powerful probe of physical 
processes occurring in the inner regions of AGN.
In particular, rapid variability is widely thought to be related to the 
central regions, and it has actually been used to constrain the physical 
properties of the central engine. 
For example, the short-time variability amplitudes of Seyfert 1
galaxies are known to be anti-correlated with the source luminosities
(Barr \& Mushotzky 1986; Nandra et al.\ 1997; Ptak et al.\ 1998).
This correlation may reflect differences in the masses of the central
supermassive black holes (SMBHs).
In addition to the short-time variabilities, the
study of long-time variabilities is also important since it may
bring out other interesting information, such as the global 
structure of the accretion disk around the central SMBH, 
or episodic events such as flares and outbursts.

Detailed studies of the long-term X-ray light curves of AGN have begun
only recently.
Markowitz \& Edelson (2001) analyzed 300-day light curves of 
Seyfert 1 galaxies in the 2$-$10~keV band.
They showed that the X-ray variability of Seyfert 1 galaxies is described
by a single, universal power-density spectrum (PDS), and that the cutoff
moves to longer timescales for sources with higher luminosity.
Soft X-ray outbursts, having an amplitude of about 2 orders of 
magnitude, were observed from NGC 5905 (Bade, Komossa, \& Dahlem 1996) 
and Zwicky 159.034 (Brandt, Pounds, \& Fink 1995).
Various possible scenarios for such outbursts have been
reviewed by Komossa \& Bade (1999).
However, long-term modulations in AGN light curves, which exceed
several years, have mostly been studied in the optical range. 
For instance, Webb (1990) reported the results of 61~years of optical
observations of 3C 120, and claimed the presence of three variability
components: a sinusoidal component with a period of 12.43 years, a linear
component, and high amplitude flares on much shorter timescales. 
Peterson et al.\ (1998) reported the spectroscopic monitoring of nine 
Seyfert 1 galaxies in the optical band, the aim of which was to determine
the size of the broad line emission regions.
In order to perform a similar analysis in the X-ray band, we need to
collect all available datasets from the various X-ray satellites.
Because the different instruments of the satellites have their own
individual properties, and cover different energy ranges, careful analysis
is required for such studies in the X-ray region.

In this paper, we study the long-term X-ray light curve of
MCG-2-58-22, covering more than 20 years, and we discuss
the potential origins of the long-term behaviors.
In previous work, Choi et al.\ (2001) analyzed the Ginga, ROSAT, 
and ASCA data for this source, and noticed flare-like events.
This motivated us to perform a thorough analysis of the long-term 
X-ray flux variations in this source. 
For this purpose, we have gathered X-ray flux measurements of
MCG-2-58-22 from the literature, as well as raw X-ray data 
from the HEASARC public archives at NASA/GSFC, and from
the SIRIUS database at ISAS\@.
The observational data we gathered include 
those of HEAO-1 (Rothschild et al.\ 1979), 
Einstein (Giacconi et al.\ 1979), EXOSAT (Taylor et al.\ 1981),
Ginga (Turner et al.\ 1989), ROSAT (Pfeffermann 1986),
ASCA (Tanaka, Inoue, \& Holt 1994), and RXTE
({\tt http://heasarc.gsfc.nasa.gov/docs/xte}).

MCG-2-58-22 is a luminous Seyfert 1 galaxy at $z=0.04732\pm0.0003$
(e.g. Huchra et al.\ 1993).
The X-ray luminosity of MCG-2-58-22, $L_X \sim 10^{44}$ erg s$^{-1}$, 
is known to be variable on a timescale of a few times $10^{3}$ seconds to 
years (Grandi et al.\ 1992; Nandra \& Pounds 1994; Choi et al.\ 2001).
Many observational characteristics are typical of Seyfert 1 galaxies:
time variability, a power-law type continuum spectrum, 
and a soft excess phenomenon (Ghosh \& Soundararajaperumal 1992; 
Nandra \& Pounds 1994; Weaver et al.\ 1995; Nandra et al.\ 1997;
George et al.\ 1998; Turner et al.\ 1999). 
The mass of the putative central SMBH of MCG-2-58-22 is estimated 
from the UV and optical observations to be a few times
$10^8$~M$_{\odot}$ (e.g., Padovani \& Rafanelli 1988; Wandel 1991).
MCG-2-58-22 has also been studied in other wavelength bands.
Mundell et al.\ (2000) observed MCG-2-58-22 using the VLBA at 8.4~GHz,
and detected the nuclear radio source, which was without any extended
structures, in the parsec-scale image; this suggests that the VLBA
radio source could be a ``bare'' Seyfert 1 nucleus (see also Weaver
et al.\ 1995).
In the optical region, this source has displayed a continuum variation
on a timescale of $\sim1$ year (de Ruiter \& Lub 1986; Whittle 1992),
as well as a very wide variability in the Balmer line profiles and
luminosities (see, e.g., Winkler et al. 1992, and references therein).

\section{DATA ANALYSIS AND LIGHT CURVE}
\subsection{Data Analysis and Flux Calibration}

To make a long-term X-ray light curve, flux data for MCG-2-58-22
have been taken from the literature; the references are listed in Table~1.
Since the instruments onboard various observatories span
different energy ranges, we convert the fluxes to the corresponding
values in the 2$-$10~keV band, which is covered by most of the
instruments. Having surveyed the reported flux data, we further
collected and analyzed the archival data of MCG-2-58-22 from the
HEASARC public archives (e.g., the EXOSAT data of 1983 November to
1984 November, the ROSAT data of 1990 November through 1993 December,
the ASCA data of 1993 May through 1997 December, and the RXTE data of
1997 December through 1999 November), and from the SIRIUS database at ISAS
in Japan (the Ginga data of 1991 June).

In the analysis of these observations, we apply the standard screening
criteria to the raw data, e.g., the rejection of possibly contaminated
data from the bright Earth, and also from regions of high particle
background, etc. Then, we calculated the X-ray fluxes in the 2$-$10 keV
band through a spectral fit to the screened data, and we list these
fluxes in Table~1.
In the spectral fit, we assumed a power-law model with a photon 
index of $\Gamma = 1.75$, and a hydrogen equivalent absorption column 
density of $N_H = 3.5 \times 10^{20}$ cm$^{-2}$ 
(see, e.g., Weaver et al.\ 1995; Piro, Matt, \& Ricci 1997).
Until now, various authors have used different models for the energy 
spectra.
Therefore, even though the best-fit parameters are available from the
literature, we use the spectral parameters stated above to maintain
consistency.
Systematic errors associated with our method are evaluated below.

\subsection{Error Estimates}

The calculated X-ray fluxes in the 2$-$10~keV band may include 
various systematic errors.    For example, there could exist
contaminating X-ray sources in the vicinity of MCG-2-58-22. 
We have checked various catalogs, and confirmed that X-ray
sources within 3 degrees of MCG-2-58-22 are dimmer than it
by at least an order of magnitude.
Thus the effect of contaminating sources can be neglected in the
present analysis.

Another possible source of error is spectral variability.
The power-law slope of MCG-2-58-22 may not be constant, and the energy  
spectra may include some structure, such as a broad iron line
or a Compton reflection hump.
The effects of spectral variability on long-term light
curves have been evaluated as follows.
We first note that the photon indices reported in the literature fall in
the range $\Gamma = 1.4-1.9$ (e.g., Turner et al. 1991;
Ghosh \& Soundararajaperumal 1992; Nandra \& Pounds 1994;
Weaver et al. 1995; George et al. 1998), 
except for the ROSAT observations.
Based on this fact, we have checked how the flux estimation is affected,
both by the change in the photon indices, and also by the structures
which may exist in the energy spectra (e.g., George et al. 1998).
As a result, it is found that differences in the spectral slope and
structures mostly affect the higher energy spectrum (say $>$5~keV),
which carries only relatively little flux.
Thus, the systematic error associated with these uncertainties 
turns out to be much smaller than 10~\%, for observations covering 
the 2$-$10~keV band.
On the other hand, since the ROSAT PSPC did not cover the 2$-$10~keV 
range, the calculated ROSAT fluxes may include relatively large
systematic errors.
In the case of the ROSAT data, it may not be appropriate to assume 
a simple power-law, because the photon indices are different between 
the 0.1$-$2~keV ROSAT band ($\Gamma \sim 2.1$) and the typical 
2$-$10~keV band ($\Gamma \sim 1.4-1.9$). 
Piro, Matt, \& Ricci (1997) attempted to use various models aimed at
simultaneously explaining both the Ginga and ROSAT data.
We utilize their models to estimate the systematic error in the flux
conversion from the PSPC count rate to the 2$-$10~keV flux.
We find that the flux conversion factors change by $\sim$20~\%
between the different models, if we take into account the observed
range of power-law slopes ($1.4-1.9$) in the 2$-$10~keV band\@.
Thus the converted ROSAT flux may include a systematic error as large
as $\sim$20~\%.

In addition to these systematic errors, we also need to consider
the cross-calibration error among the instruments.
However, accurate instrumental cross-calibration is usually difficult,
and quantitative estimations are not available for most of
the satellites.
Here, we assume 10~\% as a representative value.
Therefore, if the 2$-$10~keV band is covered by the instruments, the 
systematic error in the flux conversion is probably dominated by the
cross-calibration error among the instruments, and this may be estimated to
be $\sim$10~\%.
On the other hand, if the 2$-$10~keV band is not covered by the 
instrument, as in the case of the ROSAT PSPC, the systematic error in the
flux conversion is dominated by the uncertainty in the spectral shape,
and this may be as large as $\sim$20~\%.

\subsection{Long-Term X-ray Light Curve}

The X-ray fluxes for the long-term light curve of MCG-2-58-22
are compiled in Table~1, and they are plotted in Figure~1 along
with the above-estimated error bars.
It is readily seen that the light curve shows at least two 
distinct characteristics, i.e., a gradual and secular decrease of 
the flux, and occasional abrupt increases, or flares.
The flux steadily decreases over a duration exceeding 15 years,
from 1977 September to 1993 December.
We draw attention to the 3 highest data points, as indicated by
the arrows in the light curve, which occurred in 1983 November,
1991 June through November, and 1997 June through December, 
respectively. They are likely to have been flare events.
The data points in 1999 could indicate an onset of another
flare, or they could be a part of the 1997 flare.
The flux increased by at least a factor of $\sim\! 2-4$ during the 
flares, as compared to the long-term trend of the secular flux decrease.
The flares are not due to an underestimation of the cross-calibration
error, since they can be identified even in data from a single satellite 
(multiple flare events occurred in 1983, 1991, and 1997 for EXOSAT,
ROSAT, and ASCA \& RXTE, respectively).
Furthermore, the flares in 1991 and 1997 were observed by the imaging
instruments, ROSAT and ASCA\@. 
This strongly suggests that MCG-2-58-22 is indeed the source of 
these flares.
We also checked for whether these events could be due to artificial  
effects, such as background fluctuations, or solar X-ray contamination.
We concluded that these flares are intrinsic to the source.

The flare durations, and the intervening intervals, may give important 
information on the origin of these events.
However, it is difficult to accurately measure the true peak flux, and
thus duration, of the flares, because of the sparse data sampling.
Using the 1991 flare, whose rise is relatively well sampled, 
we estimate its likely duration to be $\sim$2 years. 
Although the observations may have missed the true peaks, 
the flare amplitudes, measured relative to the secular decreasing
trend, are rather similar, at a level of
$\sim 4\times10^{-11}$ erg~cm$^{-2}$~s$^{-1}$.
A further interesting feature is that the flares may have occurred
quasi-periodically; the time interval between the first and second flares
is $\sim$8~yrs, and it is $\sim$6~yrs between the second and third flares.
However, because the sampling is sparse, and includes long data gaps,
further observations are needed to confirm this quasi-periodicity.

\section{DISCUSSION AND CONCLUSION}

As shown in Figure~1, MCG-2-58-22 clearly shows two characteristic
variabilities: the gradual, secular decrease of the X-ray
flux, and multiple flares with a representative duration
of $\sim$2~years.
These two characteristic variabilities imply that the source
undergoes at least two distinctly different physical processes. 
Magnetic reconnection may be considered as a mechanism for the flares,
but it seems unlikely that this process could explain their duration
and repetition frequency, since the magnetic field evolution timescale
in an accretion disk is of order the dynamical time
(Romanova et al. 1998; Poutanen \& Fabian 1999).
Another possible, and more likely, origin of the variabilities are the
instabilities which could arise from the accretion disk.
Such instabilities may result in the modulation of the mass-accretion
rate, leading then to the observed flux variations.

Disk instabilities which could be appropriate for the variabilities in
MCG-2-58-22 are the viscous-thermal and the viscous instabilities.
If the disk experiences a viscous-thermal instability
caused by a sudden change in the hydrogen ionization state, we estimate 
the instability timescale to be
$t_{\rm vis-th}\sim 6\times 10^2(\alpha/0.1)^{-1} M_8^{1/3}
({\dot M}/10^{-4}M_{\odot}\ {\rm yr}^{-1})^{1/3}$~yr,
where $M_8$ is the SMBH mass in units of $10^8$~M$_{\odot}$.
On the other hand, the viscous timescale near the innermost radius 
of the standard $\alpha$-disk can be a few tens of years for a
black-hole mass of $10^8$~M$_\odot$.   
Because these timescales are much longer than the observed duration 
of the flares, the viscous-thermal and the viscous instabilities
are probably not the cause of the flares.
However, the gradual and secular decrease of the X-ray flux
has a longer timescale, and this slower variation could be caused
by these types of disk instability.

Temporal increases of the mass accretion rate could arise,
in principle, from other mechanisms. 
One entertaining possibility is the tidal disruption of stars
by the SMBH (e.g., Rees 1988; Lee \& Kim 1996; Kim et al.\ 1999). 
In this model, the frequency with which
a star passes within a distance $r$ from the SMBH can be
estimated to be
$\approx 10^{-3}~M_8^{4/3} (N_*/10^6~ {\rm pc^{-3}}) 
(\sigma/300~ {\rm km~s^{-1}}) (r/r_t)~~ {\rm yr^{-1}}$,
where $N_*$ is the number density of the stars, $\sigma$ is the virial 
velocity of the stars, and $r_t$ is the tidal radius of the SMBH\@.
Although the number density of the stars and their velocity dispersion
near the SMBH are not known, the flare events we observed from 
MCG-2-58-22 may be difficult to interpret as independent
tidal disruption events, because the event rate would then be too high.  
Instead, the resulting flares may be produced when the bound 
material from the tidally disrupted star returns to the pericenter, with
an orbital period of a few years (Rees 1988; Ulmer 1999). 
The observed peak luminosity of $\sim\! 4 \times 10^{44} ~{\rm erg~s^{-1}}$
for the flares corresponds to that from the debris of 
$\sim$0.1~M$_\odot$ being swallowed steadily with 10~\% 
efficiency over a year's duration. 
However, it is known that a star may be captured by the SMBH 
without tidal disruption, provided the SMBH is a Schwarzschild black hole
heavier than $\sim\!10^8$~M$_\odot$; this limit is comparable to
previous estimates of the SMBH mass in MCG-2-58-22.
If we consider a Kerr black hole, the tidal disruption may still
be possible, depending on the trajectory of the approaching star
(Beloborodov et al.\ 1992).
Moreover, the atmospheres of giant stars could be stripped off
before being captured by the SMBH, and this process may be the actual
cause of the flares (e.g., Nolthenius \& Katz 1982; Carter \& Luminet 1983; 
Rees 1988; Laguna et al.\ 1993; Ulmer 1999). 
Thus, the observed flares are not inconsistent with those caused by the
tidally disrupted stellar debris near the SMBH, even though the flare
properties would depend on many unknown parameters, such as: 
the type of the disrupted star, the spin of the SMBH, and the minimum
radius of the trajectory.

The long-term optical light curve of MCG-2-58-22 was obtained by
Winkler et al.\ (1992), as a result of a 4-year campaign to monitor
35 southern Seyfert galaxies.
The campaign period from 1987 through 1990 overlaps
with the Ginga observations of 1989 and 1990.
During this time, a clear variation of about 0.3 mag is reported in the
V-band. Unfortunately, however, the optical light curve does not cover
the time periods of the X-ray flares we detect, and it shows only gradual
and smooth variations. Over this time, the X-ray flux varied by about 40~\%.
This is comparable to the 0.3 mag variations in the V-band.
Thus, at least the fractional amplitude of the long-term variations
may be similar between the optical and X-ray wavelength bands.   
It is difficult to ascertain whether the optical variation is 
physically related to that in X-rays, since the optical 
observation period is too short, compared with that of the X-ray 
light curve.    

The variabilities we detected from MCG-2-58-22 may be reminiscent of
the rapid time variations seen in the galactic black 
hole candidates (GBHCs).
Phenomenologically, the presence of a characteristic timescale
is known both in GBHCs in the hard state,
and in the SMBHs (Hayashida et al.\ 1998; Markowitz \& Edelson 2001).
In both kinds of sources, the characteristic timescale corresponds
to the cutoff in the PDS, and this timescales may be scaled almost linearly 
with the black hole mass.
The canonical galactic black hole, Cyg X-1, contains about 10~M$_\odot$,
and it shows a PDS cutoff at around $\sim$10~sec.
The PDS of Cyg X-1 may be reproduced by the superposition
of shot noise impulses, whose typical duration determines
the PDS cutoff frequency (e.g.\ Negoro, Kitamoto, \& Mineshige 2001).
If we assume that the SMBH in MCG-2-58-22 has a mass of 
$10^8$~M$_\odot$ (e.g., Padovani \& Rafanelli 1998; Wandel 1991),
the characteristic timescale is about a few years.
This is just about equal to the duration of the flares we detected.
Thus the flares in MCG-2-58-22 may be analogous to the shot noise 
seen in galactic black holes in their hard state. 

In conclusion, we have detected two characteristic variabilities in
the long-term X-ray light curve of MCG-2-58-22, by analyzing archival 
data from various X-ray satellites. 
One variation is the gradual, secular decrease of the X-ray flux,
which may have a timescale of several decades, and the other is the flaring.
Although it is difficult to accurately measure the duration of, and
intervals between, the flares, due to the sparse sampling in the
observational data, a representative duration is $\lesssim $2 years,
with an intervening interval of $\lesssim\! 6-8$ years. These two distinct
timescales may be accounted for by a model with a supermassive black hole
accompanied by an unstable accretion disk; the long-term secular variation
would be expected from instabilities in the disk, while the short-term
flaring would arise from the tidal disruption of stars by the
supermassive black hole. Further observations and spectral analysis
in other wavelength bands should be explored, in order to verify
this scenario.

\acknowledgments

C.S.C. is grateful to A. Fletcher (KAO) for his careful reading of this
manuscript, as well as for useful discussions. This research has made use
of data from the HEASARC on-line service provided by the NASA GSFC,
and from the SIRIUS database of ISAS in Japan.


\begin{table}
\begin{center}
\caption{JOURNAL OF X-RAY OBSERVATIONS OF MCG$-$2-58-22}

\begin{tabular}{lcccc}  \tableline \tableline
\multicolumn{1}{l}{DATE} &
\multicolumn{1}{l}{OBSERVATORY} &
\multicolumn{1}{c}{INSTRUMENT} &
\multicolumn{1}{c}{FLUX$^a$} &
\multicolumn{1}{c}{REFERENCE$^b$} \\ \tableline

1977 September 5 $-$     & HEAO-1    & A2       & 4.0 & (1) \\
1978 March 13            &           &          &           \\
1978 March 14 $-$        & HEAO-1    & A2       & 5.0 & (1) \\
1978 September 11        &           &          &           \\
1979 May 27              & Einstein  & MPC, SSS & 4.2 & (2) \\
1979 June 2              & Einstein  & MPC, SSS & 3.8 & (2) \\
1983 November 6          & EXOSAT    & ME       & 6.7 & (3) \\
1984 November 16         & EXOSAT    & ME       & 3.0 & (3) \\
1984 November 17         & EXOSAT    & ME       & 2.9 & (3) \\
1984 November 20         & EXOSAT    & ME       & 3.1 & (3) \\
1984 November 22         & EXOSAT    & ME       & 3.2 & (3) \\
1984 November 24         & EXOSAT    & ME       & 3.2 & (3) \\
1984 November 26         & EXOSAT    & ME       & 2.9 & (3) \\
1989 June 19             & Ginga     & LAC      & 1.9 & (4) \\
1989 July 5              & Ginga     & LAC      & 2.3 & (4) \\
1989 November 5          & Ginga     & LAC      & 1.8 & (4) \\
1989 November 24         & Ginga     & LAC      & 2.8 & (4) \\
1990 November 27         & ROSAT     & PSPC-B   & 2.5 & (3) \\
1991 June $7-8$          & Ginga     & LAC      & 3.6 & (3) \\
1991 November 21         & ROSAT     & PSPC-B   & 5.8 & (3) \\
1993 May $21-25$         & ROSAT     & PSPC-B   & 1.3 & (3) \\
1993 May $24-26$         & ROSAT     & PSPC-B   & 1.5 & (3) \\
1993 May $25-26$         & ASCA      & GIS, SIS & 1.7 & (3) \\
1993 December 1          & ROSAT     & PSPC-B   & 1.0 & (3) \\
1997 June $1-2$          & ASCA      & GIS, SIS & 3.5 & (3) \\
1997 December $15-17$    & ASCA      & GIS, SIS & 3.5 & (3) \\
1997 December $15-16$    & RXTE      & PCA      & 4.0 & (3) \\ 
1999 May $28-30$         & RXTE      & PCA      & 1.8 & (3) \\
1999 June $7-9$          & RXTE      & PCA      & 1.8 & (3) \\
1999 July 30$-$ August 3 & RXTE      & PCA      & 2.8 & (3) \\
1999 November 3$-$5      & RXTE      & PCA      & 3.3 & (3) \\
\tableline

\multicolumn{5}{l}{$^a$ Calculated mean flux in units of
$10^{-11}$ erg cm$^{-2}$ s$^{-1}$, in the energy range 2$-$10 keV.}\\

\multicolumn{5}{l}{$^b$ References for the X-ray fluxes; (1) Piccinotti 
et al.\ (1982); (2) Turner et al.\ (1991); (3) this study;}\\

\multicolumn{5}{l}{(4) Nandra \& Pounds (1994).}\\ 

\end{tabular}
\end{center}
\end{table}

\clearpage

\figcaption[fig1.ps]{
Long-term X-ray light curve of MCG-2-58-22 in the energy range
2$-$10~keV\@.   Flux data are either taken from the references
indicated in Table 1, or calculated in this study using a spectral fit 
to the data.    Different symbols are used to distinguish the data from
the various observatories.   The short-dashed line, and the arrows,
indicate the gradual, secular flux decrease, and the flares, respectively.
\label{fig1}}

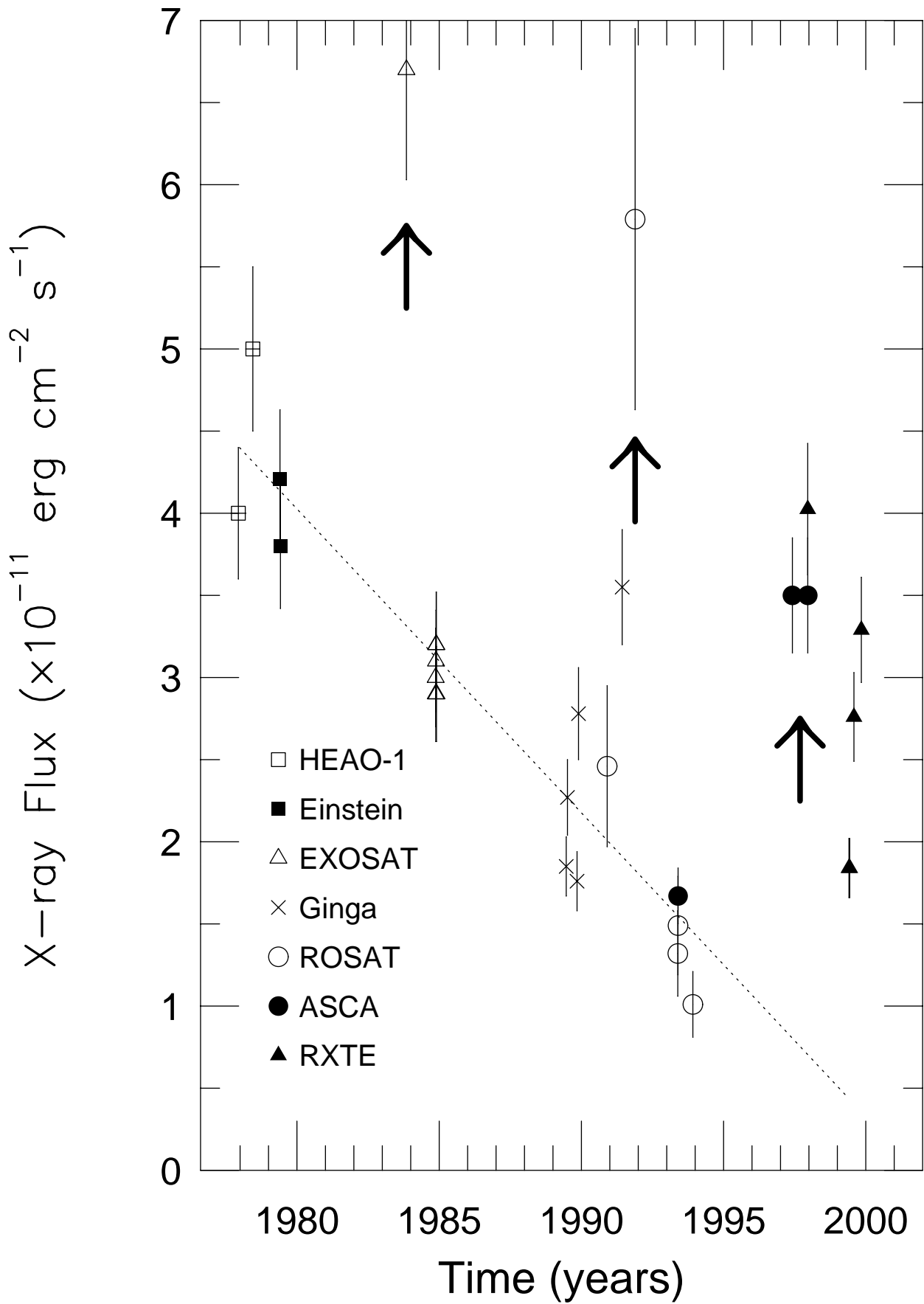
\end{document}